\documentclass[submission,copyright,creativecommons]{eptcs}

\usepackage{color}
\usepackage{amsmath, amssymb, amsthm}
\usepackage{subfig}
\usepackage{multirow}
\usepackage{tikz}
\usetikzlibrary{patterns}
\usetikzlibrary{arrows,shapes}
\tikzstyle{vertex}=[circle,minimum size=8pt,inner sep=0pt]
\tikzstyle{edge} = [draw,thick,->]
\tikzstyle{selected edge} = [draw,line width=5pt,line join=round,-,red!80]
\pgfdeclarelayer{background}
\pgfsetlayers{background,main}
\tikzstyle{sec vertex}=[circle,draw,minimum size=20pt,inner sep=0pt]
\tikzstyle{thd vertex}=[circle,draw,minimum size=8pt,inner sep=0pt]
\usepackage[ruled,vlined,linesnumbered,commentsnumbered,norelsize]{algorithm2e}

\providecommand{\event}{PDMC 2011} 

\title{Computing Optimal Cycle Mean in Parallel on CUDA%
\thanks{This work has been partially supported by the Czech Grant Agency grants
    No. 201/09/P497, 201/09/1389, 102/09/H042 and by the Slovak Research and
	Development Agency grant No. VMSP-P-0107-09.}}
\author{Ji\v{r}\'{\i} Barnat
\institute{Faculty of Informatics\\ Masaryk University\\ Brno, Czech Republic}
\email{barnat@fi.muni.cz}
\and
Petr Bauch
\institute{Faculty of Informatics\\ Masaryk University\\ Brno, Czech Republic}
\email{xbauch@fi.muni.cz}
\and
Lubo\v{s} Brim
\institute{Faculty of Informatics\\ Masaryk University\\ Brno, Czech Republic}
\email{brim@fi.muni.cz}
\and
Milan \v{C}e\v{s}ka
\institute{Faculty of Informatics\\ Masaryk University\\ Brno, Czech Republic}
\email{xceska@fi.muni.cz}
}
\def\titlerunning{Computing Optimal Cycle Mean in Parallel on CUDA}
\def\authorrunning{J. Barnat \& P. Bauch \& L. Brim \& M. \v{C}e\v{s}ka}

\newcommand{\divine}{\textsc{Di\hspace*{-1pt}V\hspace*{-1pt}inE}\xspace}
\newcommand{\BA}{\renewcommand{\baselinestretch}{1}\begin{algorithm}[!t]}
\newcommand{\EA}{\end{algorithm}\renewcommand{\baselinestretch}{1}}

\newtheorem{thm}{Theorem}[section]

\newtheorem{prop}[thm]{Proposition}

\newtheorem{define}[thm]{Definition}

\theoremstyle{remark}

\theoremstyle{definition}

\begin{document}
\maketitle

\begin{abstract}
  Computation of optimal cycle mean in a directed weighted graph has many
  applications in program analysis, performance verification in particular.  In
  this paper we propose a data-parallel algorithmic solution to the problem and
  show how the computation of optimal cycle mean can be efficiently accelerated
  by means of CUDA technology.  We show how the problem of computation of
  optimal cycle mean is decomposed into a sequence of data-parallel graph
  computation primitives and show how these primitives can be implemented and
  optimized for CUDA computation. Finally, we report a fivefold experimental
  speed up on graphs representing models of distributed systems when compared to
  best sequential algorithms.
\end{abstract}

\section{Introduction}

High quality implementation of complex computer systems, e.g. complex
embedded systems, is a major challenge today and the computer industry struggles
with how to efficiently engineer these systems. Implementations of these systems
raise complex parallelism and scheduling issues, which are in practice solved by
hand or, at best, by using emerging tools that address only a limited set of
applications with favourable properties, such as static nests of loops. One way
to tackle this challenge is to use model-driven engineering.

Model-driven engineering is a very active academic domain, driving many studies
and prototype tools, and there is an emerging industrial market with an expected
growth greater than 10\% per year (cf Forrester Consulting). Model-driven
performance analysis introduces performance analysis in the early design phases
and leading thus to design of more reliable and optimal system.

Inspections of graph cycles is one of the possible means to deal with performance
prediction. For example assume that transitions of a system are labelled with
resource consumption that the actions these transitions model impose on the system.
Then by finding the \emph{maximal cycle mean} of a graph representing the system
it is possible to approximate the worst sustainable load---the amount of resources
consumed---under which the system will operate. Unlike using Queueing
networks~\cite{Kle75} or stochastic Petri nets~\cite{Mol82} the computation of
optimal cycle mean (OCM) allows to measure the worst expectable performance over
infinite runs.

The potential use of the OCM computation for performance analysis should be more
apparent with our running example of a client/server distributed application.
Provided we are able to create a model of this application with edge-labels
representing consumption of CPU resources we intend to compute how many clients
will consume how much of the CPU of the server. Analogously, the inspection of
properties of critical cycles, and especially the computation of OCM, allows to
analyse performance of a large number of systems.

It has been shown that the system to be analysed can be modelled as a Petri
net~\cite{RH80}, Process graph~\cite{MDG98} or e.g.\ a Data flow graph~\cite{IP95}.
When an appropriate modelling formalism for the given real-world system is used,
the enumeration of a specific cycle property facilitates performance evaluation of
asynchronous systems~\cite{Bur91}, delay intensive and latency-insensitive
systems~\cite{LK06}; rate analysis and scheduling of embedded real-time
systems~\cite{MDG98}; time-separation analysis of concurrent systems~\cite{BHAB95}
and many others. The performance evaluation measures provided by OCM computation
include resource consumption (CPU, memory, bandwidth, etc.), worst expectable
latency and other measures in specific systems, e.g.\ cycle period in synchronous
systems.

Many different approaches to compute OCM have been taken so far. Dasdan et
al.~\cite{das99} gave a comprehensive list of algorithms. However, it is imperative
that the process of finding the cycles is not overly expensive since many of these
applications require the critical cycle to be found repeatedly~\cite{DG97}. To
further emphasize the necessity of having efficient algorithmic solution to this
problem we present two practical observations. The graphs representing even a small
system can be exceedingly large, containing millions of vertices. Moreover, the
number of cycles can be exponential with respect to the number of vertices thus
making the trivial inspection of all cycles in the graph impractical. Yet the
asymptotic complexity of even the best sequential algorithms is very high, which
renders the applicability of OCM-based performance analysis limited to small
systems. We intend to improve the run-time of OCM computation and consequently
the applicability of OCM-based performance analysis by employment of SIMD
parallelism.

The potential of SIMD parallelism has been recently rediscovered (first efficiently
employed in the Connection Machines~\cite{Hil87}) with the entry of affordable
graphics processing units (GPU) computation. The GPUs possess off-the-shelf
data-parallel computation capability, which was soon realized by the academic
community and has led to acceleration of various scientific computations. Among
these applications of SIMD parallelism were also examples of acceleration of graph
algorithms: several can be found in~\cite{HN07}, together with thorough
experimental evaluation on large sparse graphs.

Since the distributed computation of OCM algorithms has led to rather moderate
results~\cite{Cha06}, we attempt to utilize the massive parallelism of modern GPUs
to accelerate the OCM computation. In order to give a lucid description of the
procedure we first provide details on what limitations are imposed on the
algorithmic solution by the target architecture, i.e. the advantages and
shortcomings of modern general purpose GPUs. Subsequently, we choose among the
existing algorithms the one most appropriate for data-parallelization through
careful inspection of the relative extend of the underlying graph operations. Then
we describe the translation of this algorithm and finally conduct an experimental
study, comparing our new data-parallel version with the sequential version and also
with other sequential algorithms.

\section{Preliminaries}

In this section we briefly introduce the optimal cycle mean problem and the
corresponding terminology within the context of graph theory. As promised in the
introduction, the graph theoretical exposition of OCM is followed by an exemplary
use of the formalism as a mean for performance analysis, allowing computation of
the worst expectable resource consumption. We also describe CUDA computation
technology and we detail on which OCM algorithmic approach would be most suitable
for data-parallel implementation. The sequential version of the selected algorithm
is then thoroughly expounded.

\subsection{Related Graph Theory Definitions}

A \emph{graph} is a tuple $\mathcal{G}=(V,E)$, where $V$ is a set of
\emph{vertices} and $E\subseteq V\times V$.  As usual, cardinalities of sets
 $|V|$ and $|E|$ are denoted with $n$ and $m$, respectively. A \emph{path} in
 $\mathcal{G}$ is a non-empty sequence of edges $\pi=<e_1,\ldots,e_n>$ such that
 $\forall1\leq i\leq n:e_i=(v_i-1,v_i)\in E$. The length of path $\pi$ is denoted
as $|\pi|$ and for $\pi=<e_1,\ldots,e_n>$ equals to $n$. A path for which
 $v_0=v_n$ is called a \emph{cycle}. The set of all cycles of graph $\mathcal{G}$
is denoted with $Z_{\mathcal{G}}$.

We say that a graph is \emph{cyclic} if and only if every edge is part of some
cycle. Furthermore, we say that a graph is rooted if there is a root vertex $s$
of the graph such that all the other vertices are \emph{reachable} from it,
i.e. $\forall v\in V$ there is a path from $s$ to $v$. Should there be no such a
vertex in the graph, we \emph{augment} the graph be adding a new vertex $s\notin
V$ and edges $\{(s,v)|\forall v\in V\}$.

Let $\mathcal{G}=(V,E)$ be a graph.  A \emph{weight} function is a function
 $w:E\rightarrow\mathbb{R}$ that assigns a real weight to every edge of
 $\mathcal{G}$. We speak of weighted graph if $\mathcal{G}$ and $w$ are
given. Weight function naturally extends to paths as a sum of the weights of all
the edges on the path, i.e.  $w(\pi)\stackrel{df}{=}\sum_{i=1}^{n}w(e_i)$, where
 $\pi = <e_1,\ldots,e_n>$. In the case of augmented graphs we put $w(e)=0$ for
every newly added edge $e$.  Finally, let $\pi$ be a cycle in a graph
 $\mathcal{G}$ weighted with a weight function $w$. We define \emph{cycle mean}
of cycle $\pi$ as $\mu(\pi)\stackrel{df}{=} \frac{w(\pi)}{|\pi|}$. \emph{Minimal
cycle mean} for a given graph $\mathcal{G}$ and weight function $w$ is then
denoted with $\mu^*(\mathcal{G},w)$, where $\mu^*(\mathcal{G},w) = min \{
\mu(\pi) \mid \pi \in Z_{\mathcal{G}}\}$.  Henceforward, we will safely drop the
graph and weight function from the notation of minimal cycle mean and will refer
to minimal cycle mean simply as to \emph{optimal cycle mean} that will be
denoted by $\mu^*$.

\begin{define} \emph{OCM} \emph{problem}: For a given graph $\mathcal{G}$ and
  weight function $w$ find the minimal cycle mean.
\end{define}

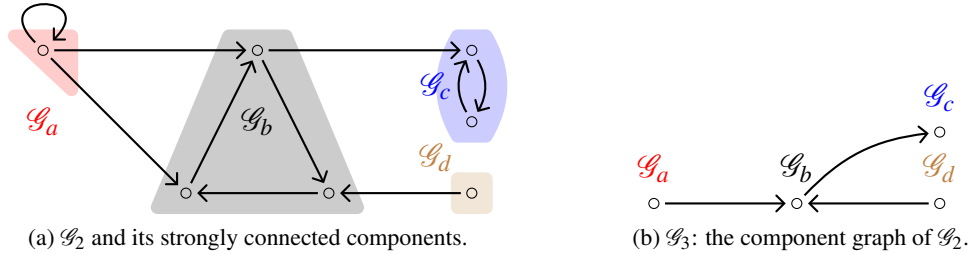
\begin{figure}
	\centering
	\subfloat[$\mathcal{G}_2$ and its strongly connected components.]{
		\label{fig:exampleSC}
		\begin{tikzpicture}[scale=1.9,auto,swap]
	\foreach \pos/\name in {{(1,2)/a}, {(2.5,2)/b}, {(4,2)/c}, {(4,1.5)/d},
		{(3,1)/e}, {(2,1)/f}, {(4,1)/g}} {
		\node[vertex] (\name) at \pos {};
		\draw \pos circle (1pt);
	}
	\begin{scope}[>=angle 90]
		\foreach \source/\dest in {a/b, b/c, g/e, b/e, f/b, e/f, a/f}
			\path[edge] (\source) -- (\dest);
		\path[edge] (c) to [bend left=30] (d);
		\path[edge] (d) to [bend left=30] (c);
		\path[edge] (a) to [in=45,out=135,loop] (a);
	\end{scope}
	\begin{pgfonlayer}{background}
		\path[draw,line width=5pt,line join=round,-,black!20,fill]
			(1.8,0.9) -- (2.3,2.1) -- (2.7,2.1) -- (3.2,0.9) -- cycle;
		\path[draw,line width=5pt,line join=round,-,red!20,fill]
			(0.8,2.1) -- (1.2,2.1) -- (1.2,1.725) -- cycle;
		\path[draw,line width=5pt,line join=round,-,blue!20,fill]
			(3.9,1.4) to [bend left=30] (3.9,2.1) -- (4.1,2.1) to
			[bend left=30] (4.1,1.4) -- cycle;
		\path[draw,line width=5pt,line join=round,-,brown!20,fill]
			(3.9,0.9) -- (3.9,1.1) -- (4.1,1.1) -- (4.1,0.9) -- cycle;
		\node[vertex] (A) at (1,1.5) {\textcolor{red}{\large{$\mathcal{G}_a$}}};
		\node[vertex] (B) at (2.5,1.5) {\textcolor{black}
			{\large{$\mathcal{G}_b$}}};
		\node[vertex] (C) at (3.75,1.75) {\textcolor{blue}
			{\large{$\mathcal{G}_c$}}};
		\node[vertex] (D) at (3.75,1.25) {\textcolor{brown}
			{\large{$\mathcal{G}_d$}}};
	\end{pgfonlayer}
\end{tikzpicture}
	}\hspace{3em}
	\subfloat[$\mathcal{G}_3$: the component graph of $\mathcal{G}_2$.]{
		\label{fig:exampleComponent}
		\begin{tikzpicture}[scale=1.9,auto,swap]
	\node[vertex] (a) at (1,1) [label=90:\textcolor{red}
		{\large{$\mathcal{G}_a$}}] {};
	\draw (1,1) circle (1pt);
	\node[vertex] (b) at (2,1) [label=90:\textcolor{black}
		{\large{$\mathcal{G}_b$}}] {};
	\draw (2,1) circle (1pt);
	\node[vertex] (c) at (3,1.5) [label=90:\textcolor{blue}
		{\large{$\mathcal{G}_c$}}] {};
	\draw (3,1.5) circle (1pt);
	\node[vertex] (d) at (3,1) [label=90:\textcolor{brown}
		{\large{$\mathcal{G}_d$}}] {};
	\draw (3,1) circle (1pt);
	\begin{scope}[>=angle 90]
		\path[edge] (a) -- (b);
		\path[edge] (b) to [bend left=20] (c);
		\path[edge] (d) -- (b);
	\end{scope}
\end{tikzpicture}
	}
	\caption{Graph with strongly connected components and the corresponding
          component graph.}
	\label{fig:examplesSCC}
\end{figure}

In order to describe further details of some OCM algorithms we introduce the
notion of parametric weight functions and strongly connected components.  Given
a weight function $w$ and a real number $\Lambda$, we define \emph{parametric
weight function} $w_{\Lambda}\stackrel{df}{=}w- \Lambda$. We say that
 $\Lambda$ is \emph{feasible} for a graph $\mathcal{G}$ if no cycle in the graph
has negative weight with respect to $w_{\Lambda}$. Directed graph $\mathcal{G}=
(V,E)$ is \emph{strongly connected} if $\forall u,v\in V$ there is a path from
 $u$ to $v$. Graph $\mathcal{G}_s=(V_s,E_s)$ is a maximal \emph{strongly
connected component} of $\mathcal{G}$ if $V_s\subseteq V$, $E_s\subseteq E$
induced on $V_s$, and $V_s$ is a maximal in $V$ such that $\mathcal{G}_s$ is
strongly connected. Graph $\mathcal{G}_c=(\mathfrak{V}_c, E_c)$ is a
\emph{component graph} of $\mathcal{G}$ if $\mathfrak{V}_c$ is the set of all
strongly connected components of $\mathcal{G}$ and
 $e=(\mathcal{G}_1,\mathcal{G}_2)$ is in $E_c\subseteq
\mathfrak{V}_c\times\mathfrak{V}_c$ if there is an edge in $\mathcal{G}$ between
a vertex from $\mathcal{G}_1$ and a vertex from $\mathcal{G}_2$. An exemplary
illustration of strongly connected components of a graph and its corresponding
component graph see Figure~\ref{fig:examplesSCC}.

\begin{prop} \label{prop:Feasible}
	$\Lambda$ is feasible $\Leftrightarrow \Lambda\leq\mu^*$.
\end{prop}

\begin{prop} \label{prop:scc}
	Minimal cycle mean of $\mathcal{G}$ is equal to the smallest of minimal cycle
	means among the strongly connected components of $\mathcal{G}$.
\end{prop}

\subsection{OCM as a Performance Analysis Measure} \label{subsec:perAn}

In order for us to be able to validate the applicability of the OCM computation
as a performance analysis measure we have extended the \divine model checking
tool with the possibility of OCM computation. Originally, the explicit state
space, parallel model checker \divine has in its core an algorithm for computation
of accepting cycle. Preserving most of the code, especially the part generating
the transition system, we only needed to perform a few changes in the modelling
language and to replace accepting cycle detection algorithms with OCM algorithms.

DVE, the modelling language of \divine, allows specification of communicating
\emph{processes} in a very intuitive fashion. From the general point of view the
processes consist of \emph{states} and we use \emph{transitions} to move from one
state to another. The communication is achieved by \emph{synchronization} of two
transitions from different processes: we require these two transition to be
executed concurrently in the resulting transition system. To allow performance
evaluation we have added specification of \emph{cost} and \emph{time} to every
transition. The cost can represent consumption of some resource of computation
(CPU utilization, memory requirements, etc.) and the addition of time allows to
compute the more general optimal cycle ratio~\cite{MDG98}.

The primary motivation behind the extension of \divine was performance analysis
of an online PDF editor that was, in the time of writing, being developed by
the Normex company. The OCM-based analysis was used with the intention to measure
the worst expectable utilization of the CPU server. For the analysis to be as
precise as possible we have devised several scenarios of the behaviour of the
clients and constructed the model as a synchronous composition of these scenarios.
Specifics of the scenarios were determined based on results of a case study among
users of a similar editor and on preliminary beta testing. The concrete findings of
our analysis will be detailed in Section~\ref{sec:experimental}.

\subsection{CUDA Computation} \label{chap:cuda}

The Compute Unified Device Architectures (CUDA)~\cite{CUDA}, developed by NVIDIA,
is a parallel programming model and a software environment providing general
purpose programming on Graphics Processing Units. At the hardware level, GPU
device is a collection of multiprocessors each consisting of eight scalar
processor cores, instruction unit, on-chip shared memory, and texture and
constant memory caches. Every core has a large set of local 32-bit registers but
no or a very small cache (L1 cache has configurable size of 16-48KB). The
multiprocessors follow the SIMD architecture, i.e. they concurrently execute the
same program instruction on different data. Communication among multiprocessors
is realized through the shared device memory that is accessible for every
processor core.

On the software side, the CUDA programming model extends the standard C/C++
programming language with a set of parallel programming supporting primitives. A
CUDA program consists of a \emph{host} code running on the CPU and a
\emph{device} code running on the GPU. The device code is structured into the so
called \emph{kernels}. A kernel executes the same scalar sequential program in
many \emph{independent data-parallel threads}.

\begin{figure}
	\centering
	\subfloat[]{\label{fig:CUDA_flow}\includegraphics[width=.35\linewidth]{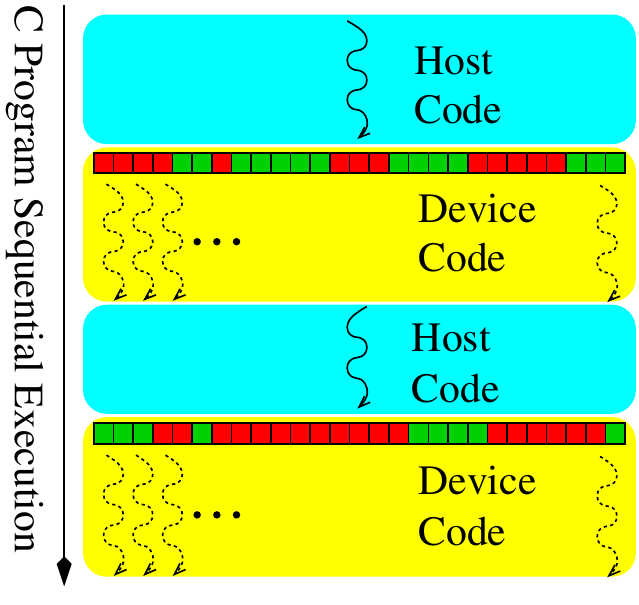}}
	\subfloat[]{\scalebox{0.7}{\label{fig:representation}
		\begin{tikzpicture}[scale=1.9,auto,swap]
	\clip (-1.25,-1.6) rectangle (6.55,0.75);
	\path[draw,thick] (0,0) -- (3,0);
	\path[draw,thick] (0,0.5) -- (3,0.5);
	\path[draw,thick] (4,0) -- (5.5,0);
	\path[draw,thick] (4,0.5) -- (5.5,0.5);

	\path[draw,thick] (0,0) -- (0,0.5);
	\path[draw,thick] (0.5,0) -- (0.5,0.5);
	\path[draw,thick] (1,0) -- (1,0.5);
	\path[draw,thick] (1.5,0) -- (1.5,0.5);
	\path[draw,thick] (2,0) -- (2,0.5);
	\path[draw,thick] (2.5,0) -- (2.5,0.5);
	\path[draw,thick] (4.5,0) -- (4.5,0.5);
	\path[draw,thick] (5,0) -- (5,0.5);
	\path[draw,thick] (5.5,0) -- (5.5,0.5);

	\node[vertex] (A) at (0.25,0.65) {\large$0$};
	\node[vertex] (B) at (0.75,0.65) {\large$1$};
	\node[vertex] (C) at (1.25,0.65) {\large$2$};
	\node[vertex] (D) at (1.75,0.65) {\large$3$};
	\node[vertex] (E) at (2.25,0.65) {\large$4$};
	\node[vertex] (F) at (5.25,0.65) {\large$|V|$};
	\node[vertex] (G) at (3.5,0.25) {\LARGE$\ldots$};

	\node[vertex] (A) at (0.25,0.25) {\large\textcolor{red}{$0$}};
	\node[vertex] (B) at (0.75,0.25) {\large\textcolor{blue}{$3$}};
	\node[vertex] (C) at (1.25,0.25) {\large\textcolor{brown}{$5$}};
	\node[vertex] (D) at (1.75,0.25) {\large\textcolor{red}{$5$}};
	\node[vertex] (E) at (2.25,0.25) {\large\textcolor{blue}{$7$}};
	\node[vertex] (F) at (4.75,0.25) {\small\textcolor{brown}{$|E|\!-\!1$}};
	\node[vertex] (F) at (5.25,0.25) {\large\textcolor{blue}{$|E|$}};

	\path[draw,thick] (-1,-1) -- (3.5,-1);
	\path[draw,thick] (-1,-0.7) -- (3.5,-0.7);
	\path[draw,thick] (4.5,-1) -- (6.0,-1);
	\path[draw,thick] (4.5,-0.7) -- (6.0,-0.7);
	\path[draw,thick,dotted] (6,-1) -- (6.5,-1);
	\path[draw,thick,dotted] (6,-0.7) -- (6.5,-0.7);

	\path[draw,thick] (-1,-1) -- (-1,-0.7);
	\path[draw,thick] (-0.5,-1) -- (-0.5,-0.7);
	\path[draw,thick] (0,-1) -- (0,-0.7);
	\path[draw,thick] (0.5,-1) -- (0.5,-0.7);
	\path[draw,thick] (1,-1) -- (1,-0.7);
	\path[draw,thick] (1.5,-1) -- (1.5,-0.7);
	\path[draw,thick] (2,-1) -- (2,-0.7);
	\path[draw,thick] (2.5,-1) -- (2.5,-0.7);
	\path[draw,thick] (3,-1) -- (3,-0.7);
	\path[draw,thick] (5,-1) -- (5,-0.7);
	\path[draw,thick] (5.5,-1) -- (5.5,-0.7);
	\path[draw,thick] (6,-1) -- (6,-0.7);
	\path[draw,thick,dotted] (6.5,-1) -- (6.5,-0.7);

	\node[vertex] (O) at (-0.75,-0.85) {\large$e_1$};
	\node[vertex] (P) at (-0.25,-0.85) {\large$e_2$};
	\node[vertex] (H) at (0.25,-0.85) {\large$e_3$};
	\node[vertex] (I) at (0.75,-0.85) {\large$e_4$};
	\node[vertex] (J) at (1.25,-0.85) {\large$e_5$};
	\node[vertex] (K) at (1.75,-0.85) {\large$e_6$};
	\node[vertex] (L) at (2.25,-0.85) {\large$e_7$};
	\node[vertex] (Q) at (2.75,-0.85) {\large$e_8$};
	\node[vertex] (M) at (5.75,-0.85) {\large$e_{|E|}$};
	\node[vertex] (N) at (4,-0.85) {\LARGE$\ldots$};

	\draw[thick] (-1.2,-1.55) rectangle (-0.3,-1.25);
	\node[vertex] (OO) at (-0.75,-1.4) {\large$(v,w,t)$};
	\path[draw,thick,densely dotted] (-1.2,-1.25) -- (-1,-1);
	\path[draw,thick,densely dotted] (-0.3,-1.25) -- (-0.5,-1);

	\begin{pgfonlayer}{background}
		\begin{scope}[>=angle 90]
			\path[edge,red] (0.25,0) -- (-0.75,-0.7);
			\path[edge,blue] (0.75,0) -- (0.75,-0.7);
			\path[edge,brown] (1.25,0) -- (1.75,-0.7);
			\path[edge,red] (1.75,0) -- (1.75,-0.7);
			\path[edge,blue] (2.25,0) -- (2.75,-0.7);
			\path[edge,brown] (4.75,0) -- (5.25,-0.7);
			\path[edge,blue] (5.25,0) -- (6.25,-0.7);
		\end{scope}
		\draw[fill,red!30] (-1,-1) rectangle (0.5,-0.7);
		\draw[fill,blue!30] (0.5,-1) rectangle (1.5,-0.7);
		\draw[fill,red!30] (1.5,-1) rectangle (2.5,-0.7);
		\draw[fill,blue!30] (2.5,-1) rectangle (3,-0.7);
		\draw[fill,brown!30] (5,-1) rectangle (6,-0.7);
		\draw[fill,blue!30] (6,-1) rectangle (6.5,-0.7);
	\end{pgfonlayer}
\end{tikzpicture}}
	}
	\caption{a) Sequential heterogeneous computation work-flow with CUDA.
			b) Adjacency list representation: a graph $G=(V,E)$ is stored as two
			arrays: $A_i$ of size $|V|+1$ and $A_t$ of size $|E|$.}
	\label{fig:complex}
\end{figure}
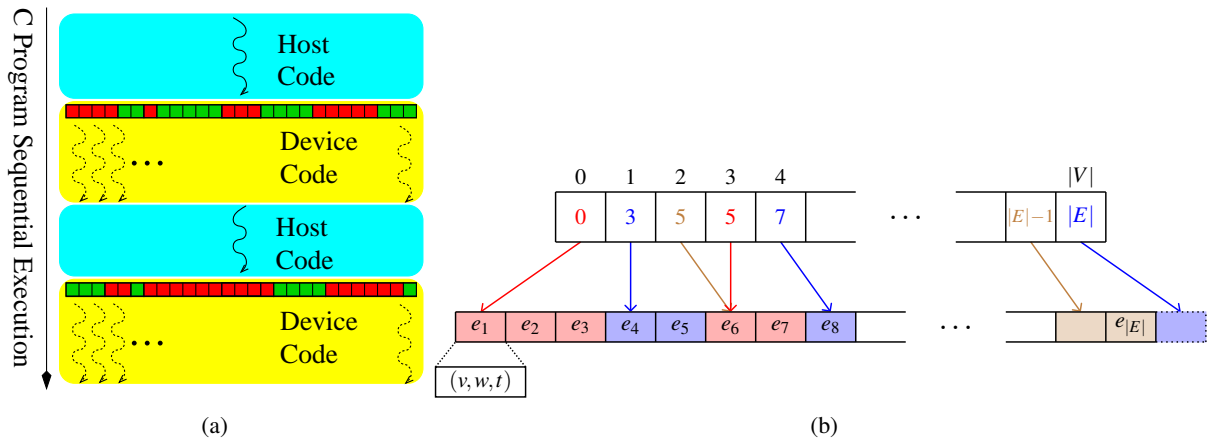

Each multiprocessor has several fine-grain hardware thread contexts, and at any
given moment, a group of threads called a \emph{warp} executes their instructions
on the multiprocessors in a lock-step manner. When several warps are scheduled on
multiprocessors, memory latencies and pipeline stalls are hidden primarily by
switching to another warp. Overall the combination of out-of-order CPU and
data-parallel processing GPU allows for heterogeneous computation as illustrated
in Figure~\ref{fig:CUDA_flow}, where sequential host code and parallel device
code are executed in turns.

Data structures used for CUDA accelerated computation must be designed with
care. First, they have to allow independent thread-local data processing so that
the CUDA hardware can fully utilize its massive parallelism. And second, they have
to be small so that the high latency device-memory access and limited
device-memory bandwidth are not large performance bottlenecks (also the regularity
of structures in the memory is of great importance). In our case, it is the
representation of graph $\mathcal{G}$ to be encoded appropriately in the first
place. Note that uncompressed matrix or dynamically linked adjacency lists violate
these requirements and as such they are inappropriate for CUDA computation.

An efficient CUDA-aware computation of a graph algorithm requires the graph to
be represented in a compact, preferably vector-like, fashion. We encode the
graph as an adjacency list that is represented as two one-dimensional arrays
 $A_t$ and $A_i$, similarly as in~\cite{HN07}. The array $A_t$ keeps target
vertices of all the edges of the graph. The target vertices stored in the array
are ordered according to the source vertices of the corresponding edges. The
second $A_i$ array then keeps an index to the first array for every vertex in
the graph. Every index points to the position of the first edge (represented as
a target vertex) emanating from the corresponding vertex. See
Figure~\ref{fig:representation}. If other data associated to a vertex are needed
by a CUDA kernel algorithm, then they are organized in vectors as well.

Most of known OCM algorithms require to access predecessors of a given vertex in
order to perform a kind of backward reachability. Storing backward edges
together with their forward versions causes additional nontrivial memory
requirements, which might be a problem as the size of CUDA memory is limited.  We
have shown how to carry out the backward reachability using only forward edges
with minor time overhead~\cite{BBBC11}. However, OCM algorithms often require
the reachability procedure to perform only on a given subgraph of the whole
graph (such that it contains a single outgoing edge for every vertex).
Unfortunately, augmenting the original procedure proposed in~\cite{BBBC11} in
order to follow the selected edges only resulted in considerable slow down in
computation. For that reason we were forced to explicitly store both forward and
backward edges of the graph.

As mentioned before the amount of memory on a CUDA device may limit the
applicability of CUDA accelerated algorithms to graphs with representation that
would fit into the memory of the GPU. Multiple CUDA-aware GPUs can be used to
effectively extend available memory~\cite{BBBC10} for the price of extensive
modification of the source code and a certain slow down. Fortunately, the memory
limitation is not that restricting in the case of OCM algorithms as the high
asymptotic complexity of individual algorithms results in practical issues
dealing with long run-times rather than a lack of memory space.

\subsection{OCM Algorithms}

Through the course of study of optimal properties of graph cycles a plethora of
algorithms emerged.
These algorithms, while not sharing similar concepts, can be divided into
several groups according to what graph property they use to find critical
cycles. Since our goal in this section is to choose one of the algorithms which
would be most suitable for data-parallel implementation we will describe all
three categories of OCM algorithms and consider their aptness for our purpose.

There are various limitations imposed on the potential algorithms should they be
even considered for SIMD acceleration. First of all, most of the data structures
used should be to a very large extend in form of vectors: stacks, queues or heaps
are not possible to be effectively data-parallelized. Then the kernels should
prevalently address the whole set of vertices (or edges): limiting the
computation to a insufficiently small subset would prevent utilization of the
computational power.  Yet the vector-wide operations are rather costly even on
many-core architectures and the complexity of algorithms is practically
measurable in their number.  Finally, there is also a non-negligible overhead of
kernel calls and thus even a very fast kernel should not be run excessively.

A strong relation can be observed between the OCM and the \emph{Shortest Path
Feasibility} (SPF) problems~\cite{DG97}. This relation should be much more
apparent once we formulate the OCM problem as a linear programming problem:
 $\mu^*$ is the optimal solution of

\begin{equation}
	\begin{array}{l}
		\mathbf{max}\ r\ \mathbf{subject\ to}\\
		\quad d(v)\leq d((u)+(w((u,v))-r)\\
		\quad \forall e=(u,v)\in E,
	\end{array}
\end{equation}

\noindent where $d$ stands for \emph{distance}, i.e. minimal-cost path from the
source node to $v$. According to the previous formulation, we can equivalently
search for the maximal parameter $r$ such that $\mathcal{G}$ with $w_r$ contains
no negative cycle.  Following Proposition~\ref{prop:Feasible} such $r$ is
exactly the optimal cycle mean. As a result, existence of an efficient
implementation of the SPF algorithm enables an efficient solution to the OCM
problem. Furthermore, any future improvements to the SPF procedure automatically
translates to improvements in this OCM problem solution.

All SPF algorithms have a common basic step: the \emph{scanning
  method}~\cite{CGGTW10}. This method assumes we maintain for every vertex $u$
its potential $\pi(u)$, parent $p(u)$, and a label $S(u)\in\{\mathit{unreached},$
 $\mathit{found},$ $\mathit{scanned}\}$. Initially, only the root vertex is
labelled $\mathit{found}$, all other vertices are $\mathit{unreached}$, and
potential of all vertices is set to zero. A single $\mathit{found}$ vertex is then
repeatedly scanned using Algorithm~\ref{alg:scanning}.

\BA
	\DontPrintSemicolon
	\SetKwInOut{Input}{Input}
	\BlankLine
	\Input{A $\mathit{found}$ vertex $u$}
	\BlankLine
	\ForEach{$e=(u,v)\in E$}{
		\If{$\pi(u)+w(e)<\pi(v)$}{
			$\pi(v),p(v),S(v)\leftarrow\pi(u)+w(e),u,\mathit{found}$\;
		}
	}
	$S(u)\leftarrow \mathit{scanned}$
	\BlankLine

	\caption{Scanning Method of SPF Algorithms}
	\label{alg:scanning}
\EA

The scanning method repeats until there are no $\mathit{found}$ vertices or until
the algorithm finishes $n$ \emph{passes}. Recall that $n=|V|$. Passes of an SPF
algorithm are defined inductively:
\begin{itemize}
	\item[$0$-th] pass is the initialization,
	\item[$i$-th] pass scans all vertices labelled $\mathit{found}$ during the
          $(i-1)$st pass.
\end{itemize}
If all $n$ passes are performed, the graph inevitably contains a negative cycle.

\subsubsection{Cycle-Based}

The arguably most straightforward application of shortest path feasibility
solution to the OCM problem is the \emph{cycle-based} approach. The idea is to
maintain an upper bound $\Lambda$ of the minimal cycle mean,
i.e. $\Lambda\geq\mu^*$, and a cycle $\mathsf{C}$ such that
 $\Lambda=\mu(\mathsf{C})$.  If $\Lambda > \mu^*$ then new better upper bound
 $\Lambda'$ of minimal mean cycle can be detected with the SPF algorithm provided
that it uses parametric weight function $w_{\Lambda}$. The newly computed upper
bound $\Lambda'$ is used in another iteration of the algorithm as
 $\Lambda$. The whole procedure repeats until no improvement of upper bound can
be found, which indicates that $\Lambda = \mu^*$.

A classical implementation of the cycle-based approach is the Howard's
algorithm~\cite{How60}, which further improves the approach by altering the SPF
procedure. At the end of each pass it checks the parent graph, induced by edges
 $(p(v),v)$, for cycles. Existence of a cycle allows to restate $\Lambda$ to a
value that sets to zero the weight of the most negative cycle. Should there be
no negative cycle the improved SPF algorithm either terminates, if all reduced
weights are non-negative, or it continues with the next pass.

The cycle-based approach seems to be fairly compatible with the SIMD
computation. Namely, the SPF subroutine is a vector-wide propagation of values,
the minimal cycle location on the parent graph is also feasible to parallelize,
and most importantly sequential experiments suggest that the algorithm typically
performs only very few passes of the underlying SPF subroutine.

\subsubsection{Binary Search}

The \emph{binary search} approach is slightly more involved. It maintains both
upper and lower bound $\Lambda_1\leq\mu*\leq\Lambda_2$ together with a cycle
 $\mathsf{C}$ such that $\mu(\mathsf{C})=\Lambda_2$. The SPF subroutine is
repeatedly called with parametric weight function $w_{\Lambda}$, where
 $\Lambda$, as the name suggests, is set to $\frac{\Lambda_1+\Lambda_2}{2}$. In
case a negative cycle $\zeta$ is found, we set $\mathsf{C}$ and $\Lambda_2$
to $\zeta$ and $\mu(\zeta)$, respectively. Since we did not use $\Lambda_2$ as a
parameter we cannot be certain of the value of the optimal cycle mean in either
of SPF answers, and thus if no negative cycle is found we set $\Lambda_1$
to $\Lambda$. The termination criterion for binary search approach is
 $\Lambda_2-\Lambda_1<\epsilon$. If $\epsilon$ is chosen sufficiently small,
 $\mathsf{C}$ will be the critical cycle.

The well-known implementation of binary search is due to Lawler~\cite{Law76}, who
also proved the run-time of his algorithm to be in $\mathcal{O}(nm\lg(W/
\epsilon))$, where $W$ is the maximal edge weight. Structurally, there is no
apparent reason why the Lawler's algorithm should be inappropriate for
data-parallelization, yet the fact that the SPF subroutine requires up to $n$
passes (and given the idea of binary search it often is necessary to carry out
all $n$ passes) renders this particular approach unusable. This hypothesis was
experimentally confirmed once we have implemented the Lawler's algorithms
and executed preliminary tests.

\subsubsection{Tree-Based}

While the previous two approaches used full SPF, the \emph{tree-based} approach
uses the shortest path feasibility subroutine only partially. Here only the lower
bound $\Lambda$ is maintained, initially small enough to guarantee that all edges
have positive weight under $w_{\Lambda}$. $\Lambda$ is progressively increased
throughout the algorithm in correctly chosen increments, until a cycle $\zeta$ is
found such that $w_{\Lambda}(\zeta)=0$. Apart from $\Lambda$ we also maintain the
shortest path tree $\mathbf{T}$, with respect to the current $w_{\Lambda}$. As we
are working with the augmented graph we may initiate $\mathbf{T}$ to consist of
the edges from $s$ to every other vertex.

The increments of $\Lambda$ must be chosen with care, otherwise minimal mean
cycle could be missed. A safe strategy is to set new value of $\Lambda$ to the
smallest $\lambda\geq\Lambda$ such that there is a different $\mathbf{T}$ for
 $w_{\lambda}$. To this end we assign to every vertex $u$ a \emph{threshold}, the
smallest $\lambda$ that would force $u$ to change parent. Finding the smallest
among all thresholds is facilitated by a priority queue.

Again this approach is unsuitable for SIMD acceleration for several reasons. The
usage of priority queues (either heaps or Fibonacci heaps) is particularly
problematic and would most likely be implemented as a simple vector, with the
minimum operation as parallel reduction. Also the span of operation updating the
shortest path tree is in many cases relatively small and would not fully utilize
the number of GPU cores. Much larger problem was found during experiments with
sequential version demonstrating that there are simply too many iteration of the
algorithm that must be carried out one after another.

\subsection{Howard's Algorithm}

Since it is the Howard's algorithm that appears to be the one most suitable for
parallelization, we will now provide its detailed description. First, we should
stress that the algorithm works on strongly connected graphs only. There exist
two approaches how to overcome this restriction. First, we can decompose the
given graph to its strongly connected components and then process the graph one
component at a time. In the sequential case we can use the Tarjan's
algorithm~\cite{Tar71} based on the depth-first traversal procedure which
outputs the list of all strongly connected component in $\mathcal{O}(n + m)$
time.  Hence there is asymptotically no difference in complexity of the
algorithm, although practically the difference can be quite substantial. The
second approach suggests to modify the underlying graph by adding a Hamiltonian
cycle $\zeta_H=<(v_0,v_1),\ldots (v_{n-1},v_0)>$ to the graph. With this
modification the graph becomes strongly connected and provided that weights of
newly added edges are sufficiently large, the optimal cycle mean of the graph
remains unchanged.

As stated in the description of the cycle-based approach, the Howard's algorithm
(see Algorithm~\ref{alg:seqHoward}, adopted from~\cite{Cha06}) extends the
shortest path feasibility algorithm. Indeed the main cycle on
lines~\ref{algLine:how1}--\ref{algLine:how2} up to line~\ref{algLine:how3} is in
fact a scanning step of the SPF algorithm. Altough the output of the cycle on
lines~\ref{algLine:how4}--\ref{algLine:how5} is not yet the shortest path tree,
since we are approaching the optimal cycle mean from top and hence there are
cycles in our shortest path graph. To remain consistent with the established
notation we will call the graph induced on edges $(v,\pi(v))$ the \emph{policy
graph}.

\BA
	\DontPrintSemicolon
	\SetKwData{D}{d}\SetKwData{Improved}{improved}\SetKwData{Val}{val}
	\SetKwData{I}{i}\SetKwData{C}{c}\SetKwData{Source}{s}\SetKwData{V}{v}
	\SetKwFunction{Succ}{Succ}\SetKwFunction{Pred}{Pred}
	\SetKwFunction{MinWeight}{MinMeanWeightCycle}
	\SetKwFunction{MinVertex}{MinVertex}
	\SetKwInOut{Input}{Input}\SetKwInOut{Output}{Output}
	\BlankLine
	\Input{A directed, strongly-connected graph $\mathcal{G}=(V,\, E,\, w), w:
		E\rightarrow\mathbb{Q}$}
	\Output{$\lambda\in\mathbb{Q}: \lambda=\mu^*(\mathcal{G})$}
	\BlankLine
	\ForEach{$v\in V$}{
		$\Val_0(v)\leftarrow0$\;
		$\pi(v)\leftarrow\mathit{nil}$\;
	}
	$\Improved\leftarrow \texttt{true}$\;
	$\I\leftarrow 0$\;
	$\lambda\leftarrow 0$\;
	\While{$\Improved$} { \label{algLine:how1}
		$\Improved\leftarrow\texttt{false}$\; \label{algLine:how11}
		\ForEach{$v\in V$} {  \label{algLine:how4}
			$\Val_{((\I+1)\mod 2)}(v)\leftarrow\min_{u\in \Succ(v)}{
				\{\Val_{(\I\mod 2)}(u)+w(v,u)-\lambda\}}$\;
			\If{$\pi(v)=\mathit{nil}\vee(\Val_{(\I\mod 2)}(\pi(v))+w(v,\pi(v))
				-\lambda > \Val_{((\I+1)\mod 2)}(v))$} {
				$\pi(v)\leftarrow u|\Val_{(\I\mod 2)}(u)+w(v,u)-\lambda=
					\Val_{((\I+1)\mod 2)}(v)$\;
				$\Improved\leftarrow\texttt{true}$\;  \label{algLine:how5}
			}
		}
		$\I\leftarrow \I+1$\; \label{algLine:how3}
		\If{$\Improved$} {
			$\C\leftarrow\MinWeight(\mathcal{G}_\pi)$\tcp*[f]{$\mathcal{G}_\pi.
				|E|=|V|, \forall v\in V: \mathit{deg}(v)=1$}
				\label{algLine:how6}\;
			$\lambda\leftarrow\mu_{\mathcal{G}(\C)}$\;
			break all other cycles than \C in $\mathcal{G}_\pi$ so that all
				vertices have path to \C\; \label{algLine:how7}
			$\Source\leftarrow\MinVertex(\C)$\; \label{algLine:how8}
			$\Val_{(\I\mod 2)}(\Source)\leftarrow 0$\; \label{algLine:how9}
			$q.\mathit{push}(\Source)$\;
		}
		\While{$\neg q.\mathit{empty}()$} { \label{algLine:how10}
			$\V\leftarrow q.\mathit{pop}()$\;
			\ForEach{$u\in \Pred(\V)$} {
				\If{$u\neq s\wedge\pi(u)=\V$} {
					$\Val_{(\I\mod 2)}(u)\leftarrow \Val_{(\I\mod 2)}(\V)+w(u,\V)-
						\lambda$\;
					$q.\mathit{push}(u)$\; \label{algLine:how2}
				}
			}
		}
	}
	\Return{$\lambda$}\;
	\BlankLine
	\caption{Howard's Algorithm}
	\label{alg:seqHoward}
\EA

Apart from the successor in the policy graph $\pi(v)$ that must exist since we
are working with strongly-connected graphs only, we also store two values
 $\mathsf{val}_0(v)$ and $\mathsf{val}_1(v)$ with every vertex. In these two
values we keep information about the current and the following parametric length
for a given vertex. In every iteration of the out-most cycle we check whether there
was a change in the policy graph, and if not we interrupt the main cycle and
return $\lambda$ as the optimal cycle mean. Each $\lambda$ that is used as a
parameter for the feasibility computation, is actually the mean weight of a
specific cycle in both the original and the policy graph. Hence, after every
iteration of the SPF algorithm part we inspect the policy graph, locate all cycles
inside it and choose the one with minimal mean weight (line~\ref{algLine:how6}).
Upon finding the minimal cycle (or after choosing one of the minimal cycles), we
modify the policy graph in such a way that every vertex has a path to the minimal
cycle (line~\ref{algLine:how7}).

Lines~\ref{algLine:how6} and~\ref{algLine:how7} would perhaps require more
detailed explanation. From a property of the policy graph (that every vertex has
exactly one outgoing edge), we know that each of its connected components consists
of one cycle and potentially several paths leading to this cycle. Finding all
cycles can thus be done in linear time simply by following the successor path,
marking all visited vertices. Next we need to rebuild the policy graph so that the
selected cycle would be the only cycle there and every vertex has a path to that
cycle. Moreover, it is required for the component of the minimal cycle to remain
unchanged, otherwise the SPF subprocedure would always detect an improvement. This
can be achieved by two consecutive backward reachabilities: one to demarcate the
component of the minimal cycle and the other one to connect also the remaining
vertices to the minimal cycle.

Subsequently, we choose one vertex (line~\ref{algLine:how8}) on the minimal
cycle and set its $\mathsf{val}$ to zero. It is necessary that we always select
the same vertex, assuming the same cycle is found minimal. After
the modification of the policy graph and selecting new $\lambda$, it is also
necessary to modify the $\mathsf{val}$ values for other vertices
accordingly. This process is started by setting the $\mathsf{val}$ of $s$ to
zero on line~\ref{algLine:how9} and carried out by the cycle on
lines~\ref{algLine:how10}--\ref{algLine:how2}, performing backward reachability
from $s$ along the edges of the policy graph. In the next iteration of the
out-most cycle we again find the policy graph (using the updated $\mathsf{val}$
values. This process is iteratively applied until two consecutive policy graphs
are found to be the same.

\section{Data-Parallel Version of Howard's Algorithm}

The actual description of our data-parallel implementation of Howard's algorithm
will be conducted in several steps. We start by proposing a high-level work
flow, where we attempt to preserve the provably correct layout. Concurrently
proposing graph primitive operations that would perform actions functionally
equivalent to those of the original algorithm, but, wherever possible,
addressing the whole vector of values at a time. CUDA-specific implementation of
these graph primitives will be detailed extensively in the following
section. Finally, we propose an extension to Howard's algorithm which prepends a
parallel decomposition to strongly connected components to the algorithm. Then
we let the algorithm perform the OCM computation on all components concurrently.

\subsection{High-Level Description}

The proposed host code of our implementation is listed as
Algorithm~\ref{alg:datHoward}. It is apparent that lines~\ref{algLine:how6}
and~\ref{algLine:how7} of Algorithm~\ref{alg:seqHoward} that rebuild the policy
graph, require much more attention in the SIMD environment as it is the place
most susceptible to inefficient processing. These two lines of CPU pseudo-code
span from line~\ref{algLine:dHow1} to line~\ref{algLine:dHow2} in our GPU
implementation. We first describe this part of the algorithm postponing the SPF
subroutine for later discussion.

\BA
	\DontPrintSemicolon
	\SetKwData{It}{it}\SetKwData{Term}{gTerminate}\SetKwData{Val}{val}
	\SetKwData{MinC}{minCycle}\SetKwData{MinI}{minIndex}\SetKwData{Mean}{mean}
	\SetKwData{PredInfo}{gPredInfo}\SetKwData{Cycles}{gCycles}
	\SetKwFunction{SpfPass}{SPFPassIter}\SetKwFunction{Elim}{elimination}
	\SetKwFunction{Ident}{cycleIdentification}\SetKwFunction{Red}{reduce}
	\SetKwFunction{Min}{min}\SetKwFunction{SetMin}{setMinCycle}
	\SetKwFunction{MarkMin}{markMinComponent}\SetKwFunction{ConG}{connectGpi}
	\SetKwFunction{GpiPre}{GpiPreprocess}\SetKwFunction{ValPro}{valuePropagate}
	\SetKwFunction{True}{true}
	\SetKwInOut{Input}{Input}\SetKwInOut{Output}{Output}
	\BlankLine
	\Input{A directed, strongly-connected graph $\mathcal{G}=(V,\, E,\, w), w:
		E\rightarrow\mathbb{Q}$}
	\Output{$\lambda\in\mathbb{Q}: \lambda=\mu^*(\mathcal{G})$}
	\BlankLine
	\While{$\True$} {
		\lIf{$\Term\leftarrow\SpfPass(\mathcal{G},\mathit{val},\lambda,
			\mathcal{G}_{\pi},\It)$}{$\mathbf{break}$}\; \label{algLine:dHow7}
		$\It++$\; \label{algLine:dHow9}
		$\GpiPre(\mathcal{G}_{\pi},\mathit{gPredInfo})$\; \label{algLine:dHow1}
		$\Elim^*(\mathcal{G}_{\pi},\mathit{gPredInfo})$\;
		$\Ident(\mathcal{G}_{\pi},\mathit{gCycles})$\; \label{algLine:dHow4}
		$\Red_{\Min}(\mathit{gCycles},\MinC)$\;
		$\lambda\leftarrow\MinC.\Mean$\;
		$\SetMin(\mathcal{G}_{\pi},\mathit{gCycles},\MinC)$\; \label{algLine:dHow5}
		$\MarkMin^*(\mathcal{G}_{\pi},\mathit{gPredInfo})$\; \label{algLine:dHow6}
		$\ConG^*(\mathcal{G},\mathcal{G}_{\pi})$\; \label{algLine:dHow2}
		$\mathit{val}[\It\&1][\MinC.\MinI]\leftarrow0$\;
		$\GpiPre(\mathcal{G}_{\pi},\mathit{gPredInfo})$\; \label{algLine:dHow3}
		$\ValPro^*(\mathcal{G}^{-1},\mathcal{G}_{\pi},\mathit{val}[\It\&1],\lambda,
			\MinC.\MinI,\mathit{gPredInfo})$\;
		 	\label{algLine:dHow8}
	}
	\Return{}\;
	\BlankLine
	\caption{Howard's Algorithm -- GPU (host code)}
	\label{alg:datHoward}
\EA

There are two calls to the \texttt{GpiPreprocess} kernel (on
lines~\ref{algLine:dHow1} and~\ref{algLine:dHow3}) and they both serve the same
purpose to gather information about predecessors in the policy graph. This step
is merely an optimization speeding up the kernels that perform backward
reachability (or its modification) on the policy graph. It would be possible to
omit this kernel for the same reason it is possible to perform backward
reachability using only forward edges~\cite{BBBC11}. Yet the speedup gained from
employing this kernel is quite considerable even though we have to call it twice
as the graph is rebuild in kernel \texttt{connectGpi}. The first call is required
because of the \texttt{elimination} and \texttt{markMinComponent} kernel, the
second call is because of \texttt{valuePropagate} kernel.

From the description of sequential Howard's algorithm we know that the policy
graph consists of \emph{weakly connected} components, each containing a cycle
and several paths leading to this cycle. In order to be able to find all cycles
in the policy graph in as few parallel steps as possible, we first apply
\texttt{elimination} to remove all vertices that do not lie on any cycle, i.e.
those that are on the paths leading to a cycle (\emph{path vertices}). This kernel
is called iteratively (every call removes the vertices with no predecessors) until
a fixpoint is found; in other words until there are no such vertices. There are
more fixpoint kernels in Algorithm~\ref{alg:datHoward} all marked with an
asterisk. The elimination allows localization of cycles and computation of their
means in a straightforward manner (line~\ref{algLine:dHow4}): we simply follow the
propagation of edges starting from any not eliminated vertex. The minimal among
them can subsequently be found by employing parallel reduction~\cite{CBZ90} with
\texttt{min} operation.

Upon finding the cycle with minimal mean (which for technical reasons has to be
agreed on by all vertices: line~\ref{algLine:dHow5}) we can actually start
rebuilding the graph. With the first backward reachability
(line~\ref{algLine:dHow6}) we undo the elimination of the path vertices within the
component of the minimal cycle, hence the second backward reachability
(line~\ref{algLine:dHow2}) is started from this component and is iteratively
applied until all vertices are connected, one breadth-first search layer at a
time.

Finally, the SPF subprocedure is easy to parallelize. Using two $\mathit{val}$
vectors, alternating the two in odd and even iterations, allows us to perform all
updates of values in a single parallel step (as there is no danger of race between
threads, see lines~\ref{algLine:dHow7},~\ref{algLine:dHow9}
and~\ref{algLine:dHow8}). Also realization of the kernel \texttt{valuePropagate}
on line~\ref{algLine:dHow8} that propagates the change of $\mathit{val}$ from the
vertex on minimal cycle with smallest index (\emph{source}), was only a minor
modification of backward reachability procedure.

\subsection{Graph Primitives and Data Structures}

We first focus on the data structures that are used during the computation. The
graph representation itself has been described in Section~\ref{chap:cuda}. In
order to keep low space profile, we store the policy graph $\mathcal{G}_{\pi}$ in
a vector of $n$ elements containing indices of the $M_c$ array that uniquely
specifies what edge leads to the successor of a given vertex. Also the first few
bits of every element are reserved to flags. For example there is a flag marking
what vertices have been removed during elimination. Cycles are stored in
 $\mathit{gCycles}$ using two 32-bit values, one for the index of its source
vertex and second for the mean weight. Finally, the vector $\mathit{gPredInfo}$
is used to store a partial information about the predecessors within one 32-bit
value. The first 16-bits are for the number of predecessors and the last 16-bits
for the \emph{local} index of the first predecessor.

While most of our kernels are only minor modifications of previously published
data-parallel graph primitives (see e.g.~\cite{BBBC11}), they are crucial for
the overall efficiency of our data-parallel implementation, and therefore, we
describe some of those modifications in detail. The \texttt{GpiPreprocess}
primitive was devised for a simple reason: there is no efficient way to propagate
along backward edges in the policy graph. Using forward edges would require
deployment of as many threads as there are unseen vertices. Searching among
backward edges those in the policy graph, on the other hand, suffers from the fact
that there are often as many edges that do not belong to the policy graph. Hence
both these approaches are approximately equally inefficient. The improvement we
have proposed first passes the whole graph $\mathcal{G}$ storing correctly the
information about predecessors within the policy graph into $\mathit{gPredInfo}$.
This preprocessing allows to virtually skip inspection of vertices with no
predecessors in the policy graph and also to jump at the first edge that belongs
to the policy graph as can be observed from the pseudo-code of
\texttt{valuePropagate} in Algorithm~\ref{alg:valProp}. There we store in the
local variable $\mathsf{counter}$ the number of predecessor of the vertex
assigned to this thread (line~\ref{algLine:valProp1}) and can skip the cycle if
 $\mathsf{counter}$ is zero. Together with the jump to the first actual
predecessor (see line~\ref{algLine:valProp2}) this improvement alone has led to
fivefold speedup of \texttt{valuePropagate} including the cost of preprocessing.

\BA
	\DontPrintSemicolon
	\SetKwData{Id}{index}\SetKwData{Tid}{threadId}\SetKwData{C}{counter}
	\SetKwData{Source}{source}\SetKwData{Pred}{pred}\SetKwData{Edge}{edge}
	\SetKwData{To}{to}
	\SetKwData{Weight}{weight}\SetKwData{Prop}{prop}\SetKwData{Fix}{fixPoint}
	\SetKwFunction{GetN}{getNum}\SetKwFunction{GetF}{getFirst}
	\SetKwFunction{True}{true}\SetKwFunction{False}{false}
	\SetKwInOut{Input}{Input}\SetKwInOut{Output}{Output}
	\BlankLine
	$\Id\leftarrow\Tid$\;
	$\Prop\leftarrow\False$\;
	$\C\leftarrow \mathit{gPredInfo}[\Id].\GetN()$\; \label{algLine:valProp1}
	$\Pred\leftarrow M_n[\Id+\mathit{gPredInfo}[\Id].\GetF()]$\;
		\label{algLine:valProp2}
	\While{$\C>0$} {
		$\Edge\leftarrow M_c[\Pred]$\;
		\If{$\Id=M_c[M_{\pi}[\Edge.\To]]$} {
			$\C--$\;
			\If{$\Edge.\To\neq\Source$} {
				$\mathit{val}[\Edge.\To]\leftarrow \mathit{val}[\Id]+\Edge.\Weight
					-\lambda$\;
				$\Prop\leftarrow\True$\;
			}
		}
		$\Pred++$\;
	}
	\If{$\Prop$} {
		$\Fix\leftarrow\False$\;
	}
	\BlankLine
	\caption{$\texttt{valuePropagate}(\mathcal{G}=(M_n,M_c),
		\mathcal{G}_{\pi}=M_{\pi},\mathit{val},\lambda,\mathsf{source},
		\mathit{gPredInfo})$}
	\label{alg:valProp}
\EA

Details on remaining kernels are as follows. \texttt{elimination} kernel is
actually the \emph{trimming} primitive (see~\cite{BBBC11}) augmented similarly
as the \texttt{valuePropagate} with the information about predecessors. A simple
while loop (it is executed only on vertices on some cycle) for identification of
cycle source and its mean is implemented in the \texttt{cycleIdentification}
kernel.  And finally the \texttt{connectGpi} performs backward reachability from
the component of the minimal cycle, and it utilizes a flag \texttt{propagate}
which marks the currently active breath-first layer. Only the threads that have a
vertex with the \texttt{propagate} flag do propagate and so less threads needs to
be dispatched.

\subsection{SCC Decomposition Extension}

There are several reasons why to prepend SCC decomposition before a CUDA
accelerated OCM algorithm.  First of all, the algorithm requires the input graph
to be strongly connected and thus we have to add the Hamiltonian cycle. Not only
is this operation costly, it also adds more edges into the graph, further
prolonging the computation. Furthermore, even though the parallel algorithms for
SCC decomposition have rather high asymptotic complexity
($\mathcal{O}(n(n+m))$), we were able to implement data-parallel SCC
decomposition which is considerable faster than the optimal sequential
algorithm~\cite{BBBC11}. And most importantly, it would allow the computation to
be executed concurrently on all SCC components, which further improves the
running time provided that the components are much smaller than the whole graph.

The technique to run a kernel on multiple \emph{regions} within a graph and to
restrict its effect to respective regions was thoroughly described
in~\cite{BBBC11} and we will thus concentrate on the parts related to
computation over decomposed graphs specific for our OCM algorithm. First of all
the Proposition~\ref{prop:scc} states that the optimal cycle mean of the whole
graph needs to be found as the minimal among all components. This observation
raises two problems, first, how to choose the minimum, and second, where to
store the component-specific $\lambda$ values during the computation (which also
has to be agreed on).

Selecting the minimal cycle mean during the computation on multiple regions is
particularly problematic as the vertices of one component are not clustered
together. It would actually require first to \emph{split}~\cite{PN09} the vector
according to the region identification and then perform \emph{segmented}
reduction~\cite{CBZ90}, both complicated and expensive operations. Fortunately,
we have observed that often very few cycles are found and consequently only a few
values are candidates for the minimum. Thus we were able to use the
\texttt{atomicCAS} operation as shown in Algorithm~\ref{alg:cas} without any
significant time expense.

\BA
	\DontPrintSemicolon
	\SetKwData{MyC}{myCycle}\SetKwData{Mean}{mean}\SetKwData{Length}{length}
	\SetKwData{Source}{source}\SetKwData{ComC}{comCycle}\SetKwData{MyR}{myRegion}
	\SetKwData{Weight}{weight}
	\SetKwFunction{AtCas}{atomicCAS}\SetKwFunction{True}{true}
	\BlankLine
	$\MyC\leftarrow(\Source,\Mean=(\Weight/\Length))$\;
	\While{$\True$} {
		$\ComC\leftarrow \mathit{cycles}[\MyR]$\;
		\lIf{$\ComC.\Mean\leq\MyC.\Mean$}{$\mathbf{break}$}\;
		$\AtCas(\&(\mathit{cycles}[\MyR]),\ComC,\MyC)$\;
	}
	\BlankLine
	\caption{Region-specific minimum voting}
	\label{alg:cas}
\EA

Also the termination needs to be modified to work on two levels. The global
termination occurs when computation is finished on all strongly connected
components. But it is also important to prevent execution of kernels on
components where we have already found the OCM. Since then less threads needs to
be deployed and less candidates compete in the minimum voting. For that purpose
we have inserted a kernel that unsets the flag \emph{work} for all vertices in
inactive components between lines~\ref{algLine:dHow7} and~\ref{algLine:dHow9} of
Algorithm~\ref{alg:datHoward}. Finally, we have optimized the overall amount of
work by unsetting the \emph{work} flag of all single vertex components prior to
the actual OCM computation.

\section{Experimental Evaluation} \label{sec:experimental}

Since the prime objective of our research was to accelerate performance analysis
based on computation of optimal cycle mean, we have compared sequential and
data-parallel algorithms mainly on models of communicating distributed systems.
The state space of all possible configurations of a given system forms a graph
with cost function (representing for example resource consumption) labelling edges
of that graph. Both modelling of the system and generation of its state space was
facilitated by the enumerative model checker DiVinE~\cite{BBCR10}, extended with
the capability to analyse performance of input models.

The sequential OCM algorithms Howard's and YTO were also implemented within
DiVinE. Thus the evaluation of our GPU Howard's algorithm is conducted by
comparing its running time against these two sequential algorithms (which are
often considered to be the fastest~\cite{DG97}). The graph representation, while
primarily targeting the vector processing architecture of GPU, is also
particularly suitable for CPU due to its cache-efficient characteristics.

All experiments were executed on a CUDA-equipped Linux workstation with an AMD
Phenom II X4 940 Processor @ 3GHz, 8 GB DDR2 @ 1066MHz RAM and NVIDIA GeForce GTX
480 GPU with 1.5 GB of GDDR5 memory. All codes were compiled with \texttt{-O3}
optimization using gcc version 4.3.2 and nvcc version 3.1 for CPU and GPU code,
respectively.

Our approach to performance analysis allows us to compute quantitative
characteristics of distributed systems where clients comply to a distinct sets
of behavioural patterns (scenarios of expected behaviour). Furthermore, we can
state how many clients of that particular scenario appear in the system and thus
we can estimate the load of a part of the system (a server for example) in an
execution based solely on the information of how many clients of what scenario
there are. As stated in Section~\ref{subsec:perAn} we have used a similar
approach in our running example for analysis of CPU utilization in an online PDF
editor (as we could for any distributed client/server application in general).
After modelling the system using the client scenarios we have computed the maximal
cycle mean in the transition system generated from the synchronous composition of
those clients. This value was equal to the worst sustainable CPU utilization
during any infinite run of the modelled system. Thus we were able to estimate the
number of clients that would be able to fully utilize the server.

In order to measure scalability of the GPU algorithm we have constructed two
distributed client/server system \emph{templates} from which a user can generate
system models by deciding on the number of client for each scenario. In the first
system, there is no server present and the actions of clients are left to be
interleaved nondeterministically. The second systems contain a server and the
clients have to communicate with the server (and only one client can access the
server at a time). The OCM of such systems then represents the average system load
inflicted on the environment.

\begin{figure}[t]
	\centering
	\includegraphics[width=\textwidth]{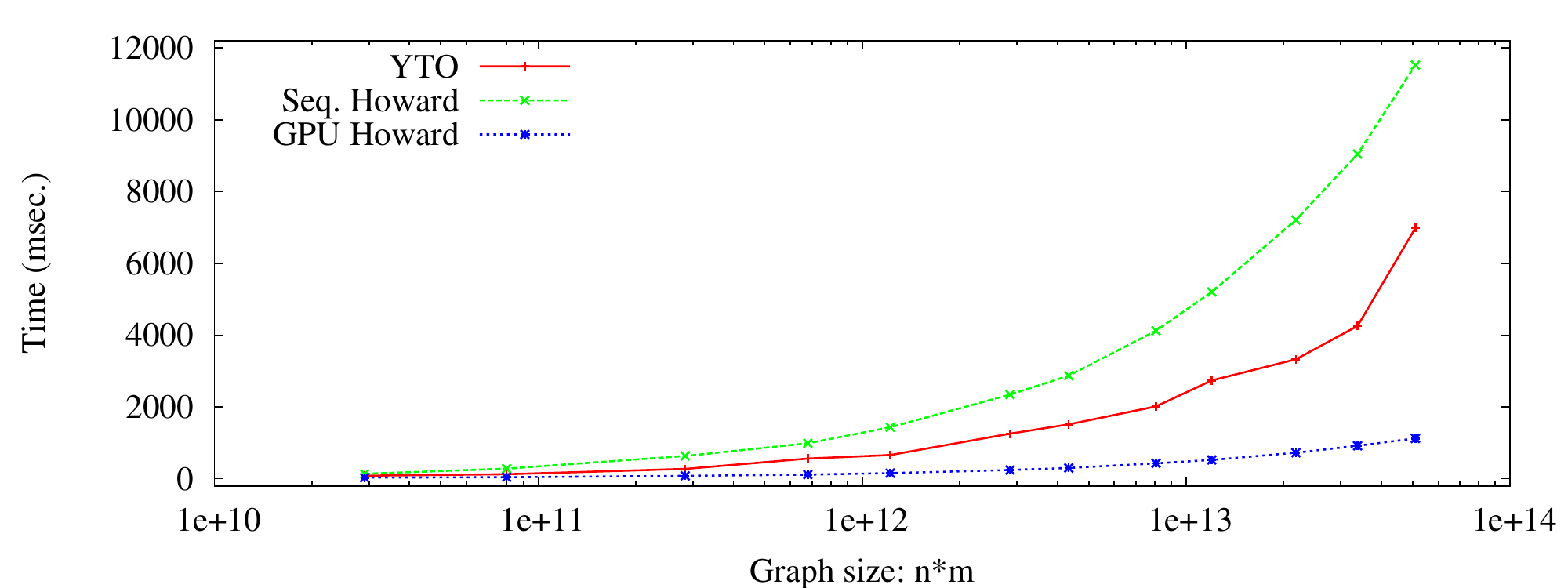}
	\caption{Plot for the server-free system.}
	\label{fig:com}
	\vspace*{-1em}
\end{figure}

\begin{figure}[t]
	\centering
	\includegraphics[width=\textwidth]{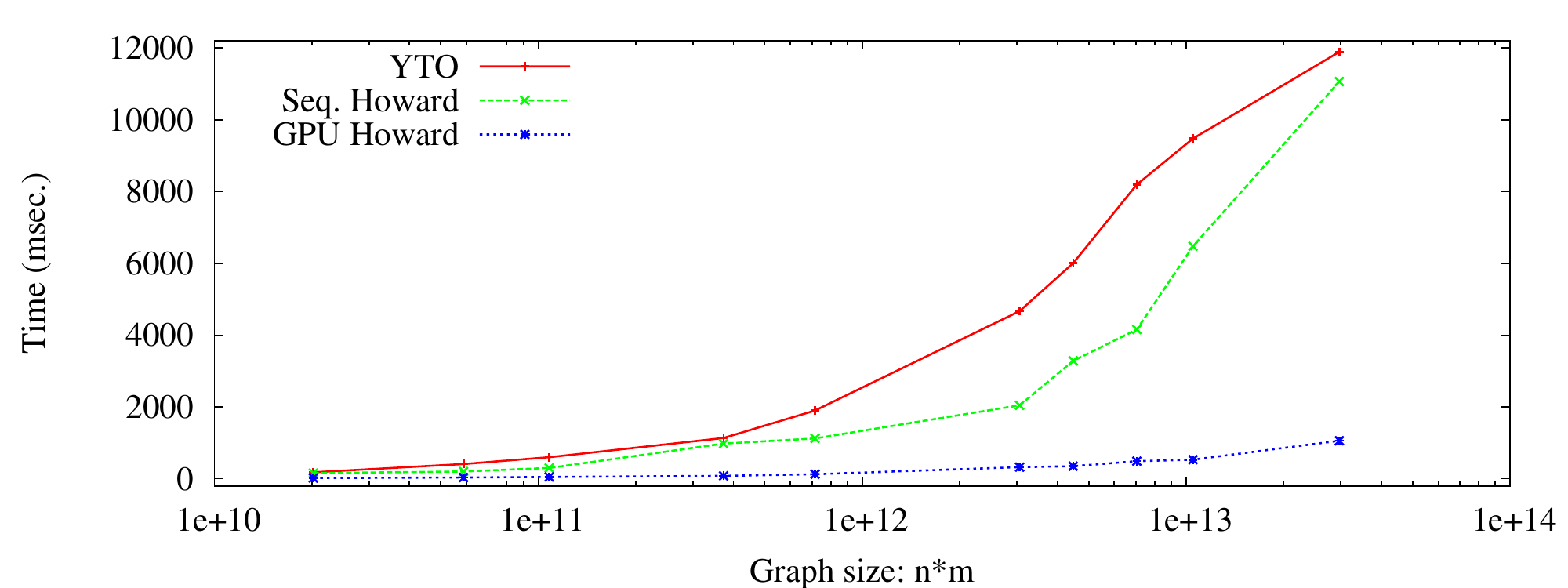}
	\caption{Plot for the system with a server.}
	\label{fig:ser}
	\vspace*{-1em}
\end{figure}

We have performed various tests on both templates to measure scalability of all
the algorithms and plotted the findings in two figures. In Figure~\ref{fig:com}
there are the results for system without a server and in Figure~\ref{fig:ser} the
results for system with a server. The $x$ axis are in logarithmic scale and
represents the size of the graph in $n*m$ as all the sequential OCM algorithms are
asymptotically in $\mathcal{O}(nm)$ or worse. Actual sizes of the analysed graphs
range from 50 thousand vertices and 400 thousand edges to approximately 2 million
vertices and 25 million edges. The plots show very clearly that while both YTO and
the sequential Howard's algorithm struggle with preserving their running times as
the graphs grow bigger, our GPU implementation is capable of performing the OCM
computation in reasonably small time. On larger instances the GPU rarely fails to
provide a fivefold speed up compared to the better of the two sequential
algorithms. It is worth noting that while the CPU algorithms were deterministic
and in different runs exhibited indistinguishable behaviour, the GPU
implementation behaves nondeterministically due to the nature of the massive
parallelism. For that reason we executed every test ten times and the result
displayed in the plots is the median of all trials.

Although primarily targeting acceleration of OCM computation for performance
analysis we feel obliged to admit that on graphs from other applications was our
data-parallel implementation much less successful. We have conducted several
experiments with US traffic network graphs, random graphs and various graphs from
the DIMACS challenge and were never able to outperform the YTO algorithm. On these
graphs the YTO performed only a very few iterations which we attach to the fact
that the OCM of these algorithms was often very close to the minimal edge weight.

\section{Conclusion}

We have proposed data-parallel acceleration of an OCM algorithm within several
consecutive steps. First we have evaluated all existing classes of OCM algorithms
with respect to their predisposition for vector processing. Subsequently, we have
described thoroughly the Howard's algorithm which was found most appropriate for
GPU acceleration and devised its data-parallel version. Specifics of the
implementation together with selected data-parallel graph primitives were then
detailed, e.g. the incorporation of SCC decomposition and the concurrent execution
of the OCM algorithm on all strongly connected components.

The primary motivation behind GPU implementation of OCM algorithms was the
acceleration of performance analysis of distributed communication systems. That
we have evaluated experimentally by constructing two scalable client/server
systems based on distinct scenarios of the clients finding our data-parallel
algorithm capable of providing performance analysis in negligible time. Although
competitiveness of the GPU algorithm on other types of graphs is questionable,
we have reported a steady fivefold speed up on performance analysis graph
against all other algorithms.

\bibliographystyle{eptcs}
\bibliography{pdmc}
\end{document}